\documentclass[preprint,12pt,nopreprintline]{elsarticle}

\usepackage{amsmath,amssymb,amsthm,mathtools}
\biboptions{numbers,sort&compress}
\usepackage[hidelinks]{hyperref}
\usepackage{enumitem}
\usepackage{microtype}
\usepackage{tikz}
\usepackage{placeins}
\usepackage{needspace}
\usepackage[labelsep=period]{caption}
\journal{Theoretical Computer Science}
\makeatletter
\let\ps@pprintTitle\ps@plain
\makeatother
\newtheorem{theorem}{Theorem}[section]

\newtheorem{proposition}[theorem]{Proposition}
\newtheorem{corollary}[theorem]{Corollary}

\theoremstyle{definition}

\newtheorem{remark}[theorem]{Remark}
\newtheorem{example}[theorem]{Example}

\newcommand{\dem}{\operatorname{dem}}
\newcommand{\demf}{\operatorname{dem}_{f}}
\newcommand{\demi}{\operatorname{demi}}
\newcommand{\demif}{\operatorname{demi}_{f}}
\newcommand{\VM}{\operatorname{VM}}
\newcommand{\EM}{\operatorname{EM}}
\newcommand{\dc}{d_c}
\newcommand{\tauf}{\tau_f}
\newcommand{\nuf}{\nu_f}

\begin{document}
\begin{frontmatter}

\title{Fractional distance-edge-monitoring in lexicographic products}
\author[1]{Xiumin Wang}
\ead{xiumin.wang@mail.nankai.edu.cn}
\author[3]{Minglei Mao}
\ead{mmldyxa@163.com}
\author[2,3]{Yaping Mao}
\ead{yapingmao@outlook.com, myp@qhnu.edu.cn}
\author[4]{Ralf Klasing\corref{cor1}}
\ead{ralf.klasing@labri.fr}

\cortext[cor1]{Corresponding author.}

\address[1]{College of Cryptology and Cyber Science, Nankai University, Tianjin 300350, China}
\address[2]{Academy of Plateau Science and Sustainability, Xining, Qinghai 810008, China}
\address[3]{School of Mathematics and Statistics, Qinghai Normal University, Xining, Qinghai 810008, China}
\address[4]{Universit\'e de Bordeaux, Bordeaux INP, CNRS, LaBRI, UMR 5800, Talence 33405, France}

\begin{abstract}
We give an exact formula for the fractional distance-edge-monitoring number of $G\circ H$, where $G$ is connected of order $m\ge2$ and $H$ has order $n\ge2$. By the previously established endpoint-monitoring property, the fractional monitoring program reduces to fractional vertex cover, whose linear-programming dual is fractional matching. Writing $\dc(H)$ for the critical difference of $H$, we obtain $\demf(G\circ H)=(mn-D(G,H))/2$, where $D(G,H)=\max_{A\in\mathcal{I}(G)}\bigl(\dc(H)|A|-n|N_G(A)|\bigr)$ and $\mathcal{I}(G)$ denotes the family of all independent sets of $G$, including the empty set, while $N_G(A)$ is the open neighborhood of $A$. This formula characterizes the products attaining $mn/2$ and gives an equality criterion for the fractional matching lower bound of Ma, Cheng, Mao and Wang. Under the stated hypotheses, the lower bound of Ma et al. is attained if and only if either factor has a fractional perfect matching or $H$ is edgeless. Two outer trees show that the factor orders and fractional matching numbers do not determine the product value.
\end{abstract}


\end{frontmatter}

\section{Introduction}

In a communication or interconnection network, vertices represent network nodes and edges represent communication links. A probe at a vertex detects a link failure when deleting the corresponding edge changes a distance measured from that vertex. More precisely, for a connected graph $G$ of order at least two, a vertex $x$ monitors an edge $e$ if $d_G(x,y)\ne d_{G-e}(x,y)$ for some vertex $y$. The minimum number of probes monitoring every edge is the distance-edge-monitoring (DEM) number $\dem(G)$. Foucaud, Kao, Klasing, Miller and Ryan~\cite{FoucaudEtAl2022} introduced this parameter and established its foundational structural and complexity properties.

Subsequent DEM research has developed structural bounds, extremal characterizations and results for specific graph families~\cite{YangKlasingMaoDeng2024,YangZhouHeMao2024,JiKlasingLiMaoZhang2024,YangYangHsiehMaoKlasing2024,JiMaoChengZhang2024,YangPseudoWheel2024}. Other studies have considered graph modifications and subdivisions~\cite{YangKlasingHeMao2024,JiSubdivision2025}, fault tolerance~\cite{YangFaultTolerance2025}, computational formulations~\cite{KarteljFilipovicKratica2025} and connections with local metric dimension~\cite{YangLocalMetricDEM2025}. Random and multiple-monitor variants have also been investigated in two preprints~\cite{YangRandomDEM2024Preprint,JiKlasingMaoNing2025Preprint}. For product networks, Li, Klasing, Mao and Ning~\cite{LiKlasingMaoNing2025} investigated integral DEM numbers for several graph products, while Shao, Li, Yang and Zhang~\cite{ShaoLiYangZhang2026} studied the strong product. A related but different notion is the monitoring edge-geodetic (MEG) number, which requires both vertices witnessing an edge to belong to the selected set. Results on MEG for graph products and other graph classes, together with studies of its algorithmic aspects, appear in~\cite{XuYangBaoZhangShao2024,FoucaudMarcilleMyintSandeepSenTaruni2025,DevDeyFoucaudNarayananSulochana2025,SunJiGao2025,BiloColliForlizziLeucci2025,FoucaudMarcilleSandeepSenTaruni2026}. A related arc-geodetic variant was studied in~\cite{DasFoucaudMarcillePavanSen2025}. These monitoring models impose different constraints and do not directly determine fractional DEM values.

Li, Mao and Klasing~\cite{LiMaoKlasing2025} introduced the fractional distance-edge-monitoring number $\demf(G)$ by assigning weights in $[0,1]$ to vertices and requiring the total weight of the monitors of each edge to be at least one. We study this fractional parameter for the \emph{lexicographic product} $G\circ H$. In this product, each vertex of the outer graph $G$ is replaced by a copy of the inner graph $H$. Copies corresponding to adjacent vertices of $G$ are completely joined. Wang, Tian and Liu~\cite{WangTianLiu1985} studied the binding number of lexicographic products, and other parameters of product networks have also been investigated~\cite{ZhangWangZhaoMaoCheng2026}. These results do not yield an exact formula for fractional distance-edge-monitoring.

For connected $G$ with at least two vertices and $H$ with at least two vertices, every edge of $G\circ H$ is monitored precisely by its two endpoints. This property and the reduction of integral DEM to vertex cover were disclosed in the Chinese patent application CN121396859A, subsequently granted as CN121396859B~\cite{WangMaoPatent2026}. Hence fractional DEM on these products reduces to fractional vertex cover and, by linear programming duality, to fractional matching. A \emph{fractional matching} of a graph $F$ is a function $w\colon E(F)\to[0,1]$ satisfying $\sum_{e\ni v}w(e)\le1$ for every $v\in V(F)$. The \emph{fractional matching number} $\nuf(F)$ is the maximum total weight $\sum_{e\in E(F)}w(e)$ over all such functions. Thus $\demf(G\circ H)=\nuf(G\circ H)$ under the stated hypotheses, and the remaining problem is to determine this fractional matching number exactly.

Ma et al.~\cite{MaChengMaoWang2022} established a lower bound for the fractional matching number of lexicographic products and proved equality whenever either factor has a fractional perfect matching. We derive an exact formula and characterize all equality cases when the outer factor is connected.

The key tool is the \emph{critical difference} of a graph $F$, defined by
\[
 \dc(F)=\max_{S\subseteq V(F)}\bigl(|S|-|N_F(S)|\bigr),
\]
where $N_F(S)=\bigcup_{v\in S}N_F(v)$ is the open neighborhood. The maximum can be attained by an independent set~\cite{Zhang1990,JardenLevitMandrescu2018}. Algorithms for critical sets were studied in~\cite{Zhang1990,Ageev1994}. The fractional Berge--Tutte formula gives $\nuf(F)=(|V(F)|-\dc(F))/2$~\cite{ScheinermanUllman1997}. We combine this identity with a decomposition of neighborhoods across fibers, following the approach used by Goddard and Swart~\cite{GoddardSwart1992} in the study of graph products. This leads to an exact expression in terms of neighborhoods of independent sets.

Our first result is an exact formula for the critical difference of $G\circ H$ for arbitrary nonempty finite graphs $G$ and $H$. We show that
\[
 \dc(G\circ H)=D(G,H)
 =\max_{A\in\mathcal I(G)}
 \bigl(\dc(H)|A|-|V(H)|\,|N_G(A)|\bigr),
\]
where $\mathcal I(G)$ denotes the family of independent sets of $G$, including the empty set. This formula shows how the neighborhoods of independent sets in the outer factor affect the result, providing information that is not captured by the preceding lower bound.

Combining this identity with the established endpoint monitoring property, we obtain the exact value
\[
 \demf(G\circ H)=\frac{mn-D(G,H)}{2}
\]
for connected $G$ of order $m\ge2$ and $H$ of order $n\ge2$.

Beyond the two sufficient conditions established by Ma et al.~\cite{MaChengMaoWang2022}, our characterization identifies an edgeless inner factor as the only additional case in which equality holds.

We also construct two outer trees with the same order and fractional matching number that yield different fractional monitoring values when combined with the same inner graph. This example shows that the fractional matching numbers of the factors alone do not determine the product value. The neighborhoods of independent sets in the outer factor are also needed.

The paper is organized as follows. Section~\ref{sec:prelim} recalls the required definitions and the reduction to fractional matching. Section~\ref{sec:exact} establishes the critical difference identity and the exact fractional monitoring formula. Section~\ref{sec:equality} characterizes equality in the lower bound of Ma et al.~\cite{MaChengMaoWang2022} and presents the example involving two trees. Section~\ref{sec:examples} gives further results for specific graph families. The final section concludes the paper.

\section{Preliminaries}\label{sec:prelim}

All graphs are finite, undirected and simple. The order of a graph is its number of vertices, and connectedness is imposed only when stated. We use standard notation as in Bondy and Murty~\cite{BondyMurty2008}. For a graph $G$ and a vertex $v\in V(G)$, let $N_G(v)=\{u\in V(G)\mid uv\in E(G)\}$ denote the open neighborhood of $v$. For $A\subseteq V(G)$, let $N_G(A)=\bigcup_{v\in A}N_G(v)$ denote its open neighborhood. This union may intersect $A$ when $A$ is not independent. For an independent set $A$, the sets $A$ and $N_G(A)$ are disjoint. We use $X\dot\cup Y$ to denote the disjoint union of two sets $X$ and $Y$. We denote by $\mathcal{I}(G)$ the family of all independent sets of $G$, including the empty set. An \emph{independent set} is a set of pairwise nonadjacent vertices, a \emph{vertex cover} is a set of vertices meeting every edge, and a \emph{matching} is a set of edges no two of which share an endpoint. The \emph{independence number} $\alpha(G)$ is the maximum cardinality of an independent set of $G$, the \emph{vertex cover number} $\tau(G)$ is the minimum cardinality of a vertex cover of $G$, and the \emph{matching number} $\nu(G)$ is the maximum cardinality of a matching of $G$. For an edge $e\in E(G)$, $G-e$ denotes the graph obtained by deleting $e$. For a vertex set $S\subseteq V(G)$, $G-S$ denotes the induced subgraph obtained by deleting the vertices in $S$. We write $i(F)$ for the number of isolated vertices of a graph $F$. We use $P_r$, $K_r$, and $K_{1,r}$ to denote the path on $r$ vertices, the complete graph on $r$ vertices, and the star with $r$ leaves, respectively. For vertices $u,v\in V(G)$ in the same component, $d_G(u,v)$ denotes the length of a shortest $u$--$v$ path; if they lie in different components, we set $d_G(u,v)=+\infty$.

\subsection{Distance-edge-monitoring and its fractional relaxation}

We follow Li, Mao and Klasing~\cite{LiMaoKlasing2025}. For an edge $e$ of a connected graph $G$, its vertex monitoring set is
\[
\VM_G(e)=\bigl\{x\in V(G)\mid
 d_G(x,y)\ne d_{G-e}(x,y)\text{ for some }y\in V(G)\bigr\}.
\]
For $x\in V(G)$, let $\EM_G(x)=\{e\in E(G)\mid x\in\VM_G(e)\}$. A set $S\subseteq V(G)$ is called a \emph{distance-edge-monitoring set} if $S\cap\VM_G(e)\ne\varnothing$ for every edge $e\in E(G)$. The minimum cardinality of such a set is denoted by $\dem(G)$. An edge set is a distance-edge-monitoring independence set if no vertex monitors two of its edges. Its maximum size is denoted by $\demi(G)$.

A function $h\colon V(G)\to[0,1]$ is a fractional distance-edge-monitoring function if $\sum_{v\in\VM_G(e)}h(v)\ge1$ for every edge $e$. Its minimum total weight is $\demf(G)$. The dual parameter $\demif(G)$ is the maximum total weight of a function $g\colon E(G)\to[0,1]$ satisfying $\sum_{e\in\EM_G(v)}g(e)\le1$ at every vertex $v$.

Let $M$ be the matrix with rows indexed by edges and columns by vertices, where $M_{e,v}=1$ when $v\in\VM_G(e)$ and $M_{e,v}=0$ otherwise. The primal and dual linear programs are
\begin{equation}\label{eq:monitoring-programs}
\begin{aligned}
\demf(G)&=\min\{\mathbf1^T\mathbf{x}\mid M\mathbf{x}\ge\mathbf1,\ \mathbf{x}\ge0\},\\
\demif(G)&=\max\{\mathbf1^T\mathbf{y}\mid M^T\mathbf{y}\le\mathbf1,\ \mathbf{y}\ge0\}.
\end{aligned}
\end{equation}
Here $\mathbf{1}$ denotes an all-ones vector of the appropriate dimension, and $\mathbf{x}$ and $\mathbf{y}$ are the nonnegative primal and dual variable vectors, indexed by $V(G)$ and $E(G)$, respectively. The upper bounds of one can be omitted without changing the optimal values of the linear programs in~\eqref{eq:monitoring-programs}. Li, Mao and Klasing~\cite{LiMaoKlasing2025} give $\dem(G)\ge\demf(G)=\demif(G)\ge\demi(G)$. We use their duality identity directly.

\subsection{Lexicographic products}

Following Hammack, Imrich and Klav\v{z}ar~\cite{HammackImrichKlavzar2011}, the \emph{lexicographic product} of graphs $G$ and $H$ is the graph $G\circ H$ with vertex set $V(G\circ H)=V(G)\times V(H)$. For $g,g'\in V(G)$ and $h,h'\in V(H)$, the pair $(g,h)(g',h')$ is an edge of $G\circ H$ if and only if $gg'\in E(G)$, or $g=g'$ and $hh'\in E(H)$.

\begin{figure}[htbp]
\centering
\begin{tikzpicture}[
  x=1cm,y=1cm,
  vertex/.style={circle,draw=black,fill=white,inner sep=0pt,minimum size=5pt},
  every node/.style={font=\small},
  fiber/.style={draw=black!55,dashed,rounded corners=5pt},
  outeredge/.style={draw=black!45,line width=0.5pt},
  inneredge/.style={draw=black,line width=1pt}
]
\draw[fiber] (-0.7,-0.65) rectangle (0.7,1.65);
\draw[fiber] (2.6,-0.65) rectangle (4.0,1.65);
\draw[fiber] (5.9,-0.65) rectangle (7.3,1.65);
\node[vertex,label=above:{$a_1$}] (a1) at (0,1.1) {};
\node[vertex,label=below:{$b_1$}] (b1) at (0,-0.1) {};
\node[vertex,label=above:{$a_2$}] (a2) at (3.3,1.1) {};
\node[vertex,label=below:{$b_2$}] (b2) at (3.3,-0.1) {};
\node[vertex,label=above:{$a_3$}] (a3) at (6.6,1.1) {};
\node[vertex,label=below:{$b_3$}] (b3) at (6.6,-0.1) {};
\draw[outeredge] (a1)--(a2) (a1)--(b2) (b1)--(a2) (b1)--(b2);
\draw[outeredge] (a2)--(a3) (a2)--(b3) (b2)--(a3) (b2)--(b3);
\draw[inneredge] (a1)--(b1) (a2)--(b2) (a3)--(b3);
\node at (0,-1.0) {$H_{g_1}$};
\node at (3.3,-1.0) {$H_{g_2}$};
\node at (6.6,-1.0) {$H_{g_3}$};
\end{tikzpicture}
\caption{The lexicographic product $P_3\circ K_2$.}
\label{fig:fibers}
\end{figure}
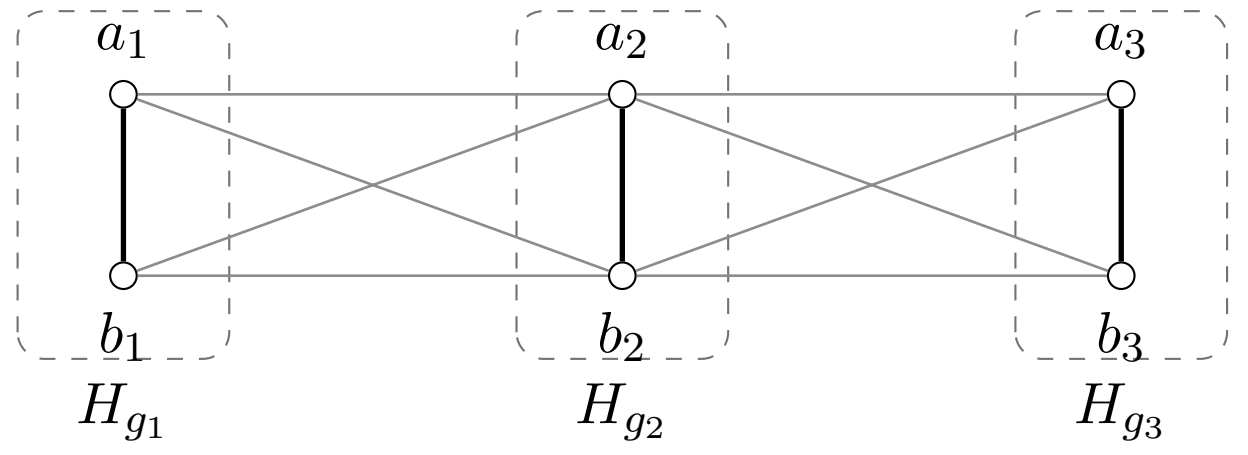

Equivalently, replace each vertex $g$ of $G$ by a copy $H_g$ of $H$. Whenever $gg'\in E(G)$, join every vertex of $H_g$ to every vertex of $H_{g'}$. The copy $H_g$, induced by $\{g\}\times V(H)$, is called the fiber over $g$. Figure~\ref{fig:fibers} illustrates this construction for $P_3\circ K_2$.
In Figure~\ref{fig:fibers}, the vertices $g_1,g_2,g_3$ occur in order on $P_3$, and $V(K_2)=\{a,b\}$. Write $a_i=(g_i,a)$ and $b_i=(g_i,b)$ for $i=1,2,3$. The dashed boxes indicate fibers, thick edges lie within fibers, and thin edges form complete bipartite graphs between consecutive fibers. Crossings without marked vertices are not additional vertices.

By Hammack, Imrich and Klav\v{z}ar~\cite{HammackImrichKlavzar2011}, when $G$ is connected of order at least two and $H$ has at least two vertices, vertices in different fibers satisfy $d_{G\circ H}((g,h),(g',h'))=d_G(g,g')$. Distinct vertices in the same fiber have distance one if their second coordinates are adjacent in $H$, and distance two otherwise. Under these hypotheses, $G\circ H$ is connected even when $H$ is disconnected.

The following endpoint-monitoring property was previously disclosed in~\cite{WangMaoPatent2026}.

\begin{proposition}\label{prop:endpoint}
Let $G$ be connected with $|V(G)|\ge2$, and let $|V(H)|\ge2$. For every edge $xy$ of $G\circ H$, its monitoring set is $\VM_{G\circ H}(xy)=\{x,y\}$.
\end{proposition}

\begin{proof}
Both endpoints monitor $xy$. Fix $z\notin\{x,y\}$ and a vertex $t$. We show that a shortest path from $z$ to $t$ avoids $xy$. The case $z=t$ is immediate. Only the edge $xy$ is deleted, while all other edges of $G\circ H$ remain unchanged.

If $xy$ lies within a fiber, vertices in different fibers admit a shortest path obtained by lifting a shortest path of $G$, using only edges between fibers. If $z,t$ lie in the same fiber, an adjacent pair uses its own edge, which is not $xy$, and a nonadjacent pair has a shortest path of length two through a neighboring fiber.

Suppose that $xy$ joins different fibers. If $z,t$ lie in the same fiber, an adjacent pair again uses its own edge. For a nonadjacent pair, choose an intermediate vertex in a neighboring fiber. Since $z$ is not an endpoint of $xy$, at most one choice uses $xy$. Another choice exists because $|V(H)|\ge2$.

Finally, suppose that $z,t$ lie in different fibers. Choose a shortest path of length $k$ between their first coordinates in $G$. If $k=1$, the edge $zt$ is a shortest path and is not $xy$. If $k\ge2$, lift this path while fixing $z,t$. The outer path has distinct vertices, so at most one step joins the two fibers containing $x,y$. If such a step occurs, at least one of its endpoints is internal to the outer path. In that fiber, choose the lifted vertex different from the corresponding endpoint of $xy$. This is possible because $|V(H)|\ge2$. Consecutive fibers are completely joined, so this replacement preserves every required adjacency and leaves the path length equal to $k$. The resulting lift is a shortest path avoiding $xy$. Hence deleting $xy$ changes no distance from $z$, which proves the assertion.
\end{proof}

Both order assumptions in Proposition~\ref{prop:endpoint} matter. If $G=P_3$ and $H=K_1$, or if $G=K_1$ and $H=P_3$, the product is $P_3$. Label this path $p,q,r$ in order. The vertex $r$ monitors the edge $pq$, since deleting $pq$ changes its distance to $p$ from two to $+\infty$. Thus the endpoint monitoring property can fail when either factor has only one vertex.

\Needspace{8\baselineskip}
For a graph $G$, the fractional vertex cover and fractional matching numbers are
\begin{equation}\label{eq:cover-matching}
\begin{aligned}
\tauf(G)&=\min\left\{\sum_{v\in V(G)}x_v\ \middle|\
 x_u+x_v\ge1\ (uv\in E(G)),\ x\ge0\right\},\\
\nuf(G)&=\max\left\{\sum_{e\in E(G)}y_e\ \middle|\
 \sum_{e\ni v}y_e\le1\ (v\in V(G)),\ y\ge0\right\}.
\end{aligned}
\end{equation}
These programs are dual, so $\tauf(G)=\nuf(G)$~\cite{ScheinermanUllman1997}. A fractional matching is perfect when its incident weights sum to one at every vertex. Equivalently, its total weight is $|V(G)|/2$.

\Needspace{10\baselineskip}
\begin{proposition}\label{prop:reduction}
Let $G$ be connected of order $m\ge2$, and let $H$ have order $n\ge2$. Then $\dem(G\circ H)=\tau(G\circ H)$ and $\demi(G\circ H)=\nu(G\circ H)$. Moreover, $\demf(G\circ H)=\demif(G\circ H)=\tauf(G\circ H)=\nuf(G\circ H)$.
\end{proposition}

\begin{proof}
By Proposition~\ref{prop:endpoint}, a set monitors all edges exactly when it is a vertex cover, and a set of edges satisfies the monitoring independence condition exactly when it is a matching. The fractional constraints in~\eqref{eq:monitoring-programs} become $x_u+x_v\ge1$ for every edge $uv\in E(G\circ H)$ and $\sum_{e\ni v}y_e\le1$ for every vertex $v\in V(G\circ H)$. These are precisely the programs in~\eqref{eq:cover-matching}.
\end{proof}

The classical identity $\alpha(G\circ H)=\alpha(G)\alpha(H)$ is due to Geller and Stahl~\cite{GellerStahl1975}. Together with $\tau(G)=|V(G)|-\alpha(G)$, it gives the integral formula $\dem(G\circ H)=mn-\alpha(G)\alpha(H)$.

\FloatBarrier
\section{The exact fractional formula}\label{sec:exact}

Recall that the critical difference of a graph $G$ is $\dc(G)=\max_{S\subseteq V(G)}(|S|-|N_G(S)|)$. The empty set is included, so $\dc(G)\ge0$. A classical result on critical independent sets~\cite{Zhang1990,JardenLevitMandrescu2018} gives
\begin{equation}\label{eq:critical}
\dc(G)=\max_{I\in\mathcal{I}(G)}
\bigl(|I|-|N_G(I)|\bigr).
\end{equation}
Thus an independent set always attains the maximum. An independent set $I$ satisfying $|I|-|N_G(I)|=\dc(G)$ is called a critical independent set of $G$. For a general vertex set $S$, the open neighborhood $N_G(S)$ may intersect $S$; for an independent set, the two sets are disjoint.

The fractional Berge--Tutte formula~\cite{ScheinermanUllman1997} states that
\begin{equation}\label{eq:berge}
\nuf(G)=\frac12\left(|V(G)|-
\max_{S\subseteq V(G)}\bigl(i(G-S)-|S|\bigr)\right),
\end{equation}
where $i(G-S)$ counts isolated vertices. The maximum in~\eqref{eq:berge} equals $\dc(G)$. To see this, take any independent set $I$ and set $S=N_G(I)$. Every vertex of $I$ is isolated in $G-S$, so $i(G-S)-|S|\ge |I|-|N_G(I)|$. By~\eqref{eq:critical}, the maximum is therefore at least $\dc(G)$. Conversely, for any $S\subseteq V(G)$, let $I$ be the set of all isolated vertices of $G-S$. Then $I$ is independent and $N_G(I)\subseteq S$, giving $i(G-S)-|S|=|I|-|S|\le |I|-|N_G(I)|\le\dc(G)$. These two inequalities establish the identity. Consequently, $\tauf(G)=\nuf(G)=(|V(G)|-\dc(G))/2$. In particular, $G$ has a fractional perfect matching if and only if $\dc(G)=0$.

\begin{theorem}\label{thm:critical-product}
Let $G$ and $H$ be graphs with $|V(G)|\ge1$ and $|V(H)|=n\ge1$. Define
\begin{equation}\label{eq:D}
D(G,H)=\max_{A\in\mathcal{I}(G)}
\bigl(\dc(H)|A|-n|N_G(A)|\bigr).
\end{equation}
Then $\dc(G\circ H)=D(G,H)$. As defined above, $\mathcal{I}(G)$ includes the empty set.
\end{theorem}

\begin{proof}
Write $d=\dc(H)$. Let $S$ be an independent set of $G\circ H$. For each $g\in V(G)$, define $S_g=\{h\in V(H)\mid(g,h)\in S\}$, and let $A=\{g\in V(G)\mid S_g\ne\varnothing\}$. Adjacent fibers are completely joined, so $A$ is independent in $G$ and each $S_g$ is independent in $H$.

Every fiber indexed by $N_G(A)$ lies entirely in $N_{G\circ H}(S)$. Within a fiber indexed by $g\in A$, the neighbors of $S$ correspond exactly to $N_H(S_g)$. No other fiber contributes a neighbor. Hence the neighborhood decomposes as
\begin{equation}\label{eq:neighborhood}
N_{G\circ H}(S)=\bigl(N_G(A)\times V(H)\bigr)
\,\dot\cup\,
\bigcup_{g\in A}\bigl(\{g\}\times N_H(S_g)\bigr).
\end{equation}
The union is disjoint since $A$ is independent and hence $A\cap N_G(A)=\varnothing$, while the fibers indexed by distinct vertices of $G$ are disjoint. This fiberwise decomposition is related to the product-neighborhood analysis of Goddard and Swart~\cite{GoddardSwart1992}.

We now count the vertices in the two sets appearing in the difference. By the definition of $S_g$, the set $S$ contains $|S_g|$ vertices in each fiber indexed by $g\in A$ and no vertices in any other fiber. Therefore
\[
|S|=\sum_{g\in A}|S_g|.
\]
In~\eqref{eq:neighborhood}, every fiber over $N_G(A)$ contributes exactly $n$ vertices. In addition, the fiber over each $g\in A$ contributes exactly $|N_H(S_g)|$ vertices to the neighborhood. Because these contributions are disjoint,
\[
|N_{G\circ H}(S)|
=n|N_G(A)|+\sum_{g\in A}|N_H(S_g)|.
\]
Subtracting the second identity from the first and regrouping the terms yields
\begin{align*}
|S|-|N_{G\circ H}(S)|
&=\sum_{g\in A}|S_g|-n|N_G(A)|
   -\sum_{g\in A}|N_H(S_g)|\\
&=\sum_{g\in A}\bigl(|S_g|-|N_H(S_g)|\bigr)
   -n|N_G(A)|\\
&\le\sum_{g\in A}d-n|N_G(A)|\\
&=d|A|-n|N_G(A)|\\
&\le D(G,H).
\end{align*}
For the first inequality, recall that $d=\dc(H)$ is the maximum of $|T|-|N_H(T)|$ over all subsets $T\subseteq V(H)$. Thus $|S_g|-|N_H(S_g)|\le d$ holds for each $g\in A$, and summing these $|A|$ inequalities yields the claimed bound. The final inequality follows because $A$ is independent in $G$ and is therefore one of the sets over which the maximum in~\eqref{eq:D} is taken. Finally, since $S$ was an arbitrary independent set of $G\circ H$, the critical-independent-set identity~\eqref{eq:critical}, applied to $G\circ H$, proves $\dc(G\circ H)\le D(G,H)$.

For the reverse inequality, if $D(G,H)=0$, the empty set suffices. Suppose that $D(G,H)>0$. If $d=0$, then every term in~\eqref{eq:D} is $-n|N_G(A)|\le0$, while the empty set contributes zero. Consequently, $D(G,H)=0$, which is a contradiction. Hence $d>0$. By~\eqref{eq:critical}, there is a nonempty independent set $J$ of $H$ satisfying $|J|-|N_H(J)|=d$. Choose an independent set $A$ attaining the maximum in~\eqref{eq:D}, which must be nonempty since $D(G,H)>0$. The set $S=A\times J$ is independent in $G\circ H$. Equation~\eqref{eq:neighborhood} now gives $|S|-|N_{G\circ H}(S)|=d|A|-n|N_G(A)|=D(G,H)$. Therefore $\dc(G\circ H)\ge D(G,H)$, proving the result.
\end{proof}

Theorem~\ref{thm:critical-product} does not require connectedness. The additional hypotheses in the following monitoring theorem are needed for Proposition~\ref{prop:endpoint}.

\begin{theorem}\label{thm:main}
Let $G$ be connected of order $m\ge2$, and let $H$ be an arbitrary graph of order $n\ge2$. With $D(G,H)$ defined in~\eqref{eq:D}, the exact value is $\demf(G\circ H)=(mn-D(G,H))/2$. In particular, $\demf(G\circ H)=mn/2$ if and only if $\dc(H)|A|\le n|N_G(A)|$ for every $A\in\mathcal{I}(G)$.
\end{theorem}

\begin{proof}
Proposition~\ref{prop:reduction} identifies $\demf(G\circ H)$ with $\nuf(G\circ H)$. The identity following~\eqref{eq:berge} and Theorem~\ref{thm:critical-product} give $\nuf(G\circ H)=(mn-D(G,H))/2$. The value is $mn/2$ exactly when $D(G,H)=0$. Since the empty set contributes zero to~\eqref{eq:D}, this is equivalent to the stated inequalities.
\end{proof}

The inner factor enters this formula only through its order and its critical difference, or equivalently its fractional matching number. For the outer factor, the formula retains the neighborhood size of each independent set. This asymmetry will be visible in Example~\ref{ex:separation}.

The exact formula yields the following useful consequence.

\begin{corollary}\label{cor:perfect}
Under the hypotheses of Theorem~\ref{thm:main}, if either $G$ or $H$ has a fractional perfect matching, then $\demf(G\circ H)=mn/2$.
\end{corollary}

\begin{proof}
By the identity following~\eqref{eq:berge}, a graph has a fractional perfect matching if and only if its critical difference is zero. Put $b=\dc(H)$, so $0\le b\le n$. If $H$ has a fractional perfect matching, then $b=0$, and every term in~\eqref{eq:D} is $-n|N_G(A)|\le0$. If $G$ has a fractional perfect matching, then $\dc(G)=0$. By~\eqref{eq:critical}, every independent set $A$ of $G$ satisfies $|A|\le|N_G(A)|$, and hence
\[
b|A|-n|N_G(A)|\le b|N_G(A)|-n|N_G(A)|\le0.
\]
In either case, the empty set contributes zero to~\eqref{eq:D}, so $D(G,H)=0$. The conclusion follows from Theorem~\ref{thm:main}.
\end{proof}

In particular, complete graphs of order at least two have fractional perfect matchings, so $\demf(K_m\circ H)=mn/2$ and $\demf(G\circ K_n)=mn/2$ whenever the hypotheses of Theorem~\ref{thm:main} hold.

Write $\overline{K_n}$ for the edgeless graph of order $n$.

\begin{corollary}\label{cor:edgeless}
For every connected graph $G$ of order $m\ge2$ and every integer $n\ge2$, we have $\demf(G\circ\overline{K_n})=n\nuf(G)$.
\end{corollary}

\begin{proof}
Since $\dc(\overline{K_n})=n$, equations~\eqref{eq:critical} and~\eqref{eq:D} give $D(G,\overline{K_n})=n\dc(G)$. The conclusion follows from Theorem~\ref{thm:main} and $\nuf(G)=(m-\dc(G))/2$.
\end{proof}

\begin{corollary}\label{cor:universal}
For a connected graph $G$ of order $m\ge2$, the following are equivalent.
\begin{enumerate}[label=(\roman*),itemsep=0.15em]
\item $G$ has a fractional perfect matching.
\item For every graph $H$ of order $n\ge2$, we have $\demf(G\circ H)=mn/2$.
\item $\demf(G\circ\overline{K_2})=m$.
\end{enumerate}
\end{corollary}

\begin{proof}
Assume (i). By Corollary~\ref{cor:perfect}, every graph $H$ of order at least two satisfies $\demf(G\circ H)=mn/2$, establishing (ii). In particular, choosing $H=\overline{K_2}$ yields $\demf(G\circ\overline{K_2})=m$, so (ii) implies (iii). Conversely, assume (iii). Corollary~\ref{cor:edgeless} gives $\demf(G\circ\overline{K_2})=2\nuf(G)$, and therefore $\nuf(G)=m/2$. By the definition of a fractional perfect matching, this means that $G$ has a fractional perfect matching, which is (i).
\end{proof}

\section{Equality in the lower bound of Ma et al.}\label{sec:equality}

Ma, Cheng, Mao and Wang~\cite{MaChengMaoWang2022} proved the following lower bound in Corollary~3.2 of their paper for lexicographic products of graphs of orders $m$ and $n$:
\begin{equation}\label{eq:ma}
\nuf(G\circ H)\ge (m-\nuf(G))\nuf(H)
+(n-\nuf(H))\nuf(G).
\end{equation}
They also proved that equality holds whenever either factor has a fractional perfect matching. Under the additional assumption that $G$ is connected, we establish a necessary and sufficient condition for equality, including the edgeless inner factor as a third case.

\begin{theorem}\label{thm:equality}
Let $G$ be connected of order $m\ge2$, and let $H$ have order $n\ge2$. Equality holds in~\eqref{eq:ma} if and only if $G$ has a fractional perfect matching, $H$ has a fractional perfect matching, or $H$ is edgeless.
\end{theorem}

\begin{proof}
Set $a=\dc(G)$ and $b=\dc(H)$. Since the empty set is admissible in the definition of the critical difference and $|S|-|N_F(S)|\le |S|\le |V(F)|$ for every $S\subseteq V(F)$, we have $0\le a\le m$ and $0\le b\le n$. Using $\nuf(G)=(m-a)/2$ and $\nuf(H)=(n-b)/2$, the right-hand side of~\eqref{eq:ma} becomes
\begin{align*}
&\left(m-\frac{m-a}{2}\right)\frac{n-b}{2}
 +\left(n-\frac{n-b}{2}\right)\frac{m-a}{2}\\
&\qquad=\frac{(m+a)(n-b)+(n+b)(m-a)}{4}
=\frac{mn-ab}{2}.
\end{align*}
By Theorem~\ref{thm:critical-product} and the fractional Berge--Tutte identity, $\nuf(G\circ H)=(mn-D(G,H))/2$. Consequently, the bound~\eqref{eq:ma} is equivalent to $D(G,H)\le ab$, and equality holds precisely when $D(G,H)=ab$.

For every independent set $A$ of $G$, equation~\eqref{eq:critical} gives $|A|-|N_G(A)|\le a$. Hence
\begin{align*}
b|A|-n|N_G(A)|
&=b\bigl(|A|-|N_G(A)|\bigr)-(n-b)|N_G(A)|\\
&\le ab-(n-b)|N_G(A)|\le ab.
\end{align*}
Thus $D(G,H)\le ab$, and equality in~\eqref{eq:ma} requires equality in these estimates.

If $a=0$ or $b=0$, Corollary~\ref{cor:perfect} and Theorem~\ref{thm:main} give $D(G,H)=0=ab$. If $b=n$, equation~\eqref{eq:critical} gives $D(G,H)=n\dc(G)=ab$. The condition $b=n$ is equivalent to $H$ being edgeless. Indeed, difference $n$ can occur only for the full vertex set with empty neighborhood.

It remains to exclude all other cases. Suppose $a>0$ and $0<b<n$. Every nonempty independent set $A$ of connected $G$ has $|N_G(A)|\ge1$. The preceding calculation therefore gives $b|A|-n|N_G(A)|\le ab-(n-b)<ab$. The empty set also gives a value strictly smaller than $ab$, since $ab>0$. The maximum in~\eqref{eq:D} is over a finite family, so $D(G,H)<ab$. Thus the bound is strict in every remaining case.
\end{proof}

The connectedness condition is relevant to the strictness argument. For example, if $G$ is edgeless of order $m$, then $G\circ H$ is the disjoint union of $m$ copies of $H$ and equality in~\eqref{eq:ma} holds for every $H$. This case lies outside Theorem~\ref{thm:equality}.

\begin{example}\label{ex:strict}
For $G=H=P_3$, both factors have fractional matching number one and critical difference one. Every independent set of $P_3$ satisfies $|A|\le2|N_{P_3}(A)|$, so $D(P_3,P_3)=0$. Consequently, $\nuf(P_3\circ P_3)=9/2$, whereas the right-hand side of~\eqref{eq:ma} equals $4$.
\end{example}

\begin{example}\label{ex:separation}
Let $G_1$ be obtained by subdividing one edge of $K_{1,5}$. Write its vertex set as $\{c,t,z,x_1,x_2,x_3,x_4\}$, with edges $ct$, $tz$ and $cx_i$ for $1\le i\le4$. Let $G_2$ be the double star with adjacent centers $u,v$, three leaves attached to $u$ and two leaves attached to $v$. Take $H=P_3$. The two outer trees are shown in Figure~\ref{fig:separation}. In $G_1$, the filled vertices in the figure mark the independent set $A=\{x_1,x_2,x_3,x_4\}$, whose neighborhood is $\{c\}$.

\begin{figure}[htbp]
\centering
\begin{minipage}[t]{0.48\textwidth}
\centering
\begin{tikzpicture}[
  x=1cm,y=1cm,
  vertex/.style={circle,draw=black,fill=white,inner sep=0pt,minimum size=5pt},
  selected/.style={vertex,fill=black,minimum size=6pt},
  every node/.style={font=\small},
  every path/.style={line width=0.7pt}
]
\node[vertex,label=below:{$c$}] (c) at (0,0) {};
\node[vertex,label=below:{$t$}] (t) at (1.3,0) {};
\node[vertex,label=right:{$z$}] (z) at (2.6,0) {};
\node[selected,label=left:{$x_1$}] (x1) at (-1.15,1.1) {};
\node[selected,label=left:{$x_2$}] (x2) at (-1.5,0.35) {};
\node[selected,label=left:{$x_3$}] (x3) at (-1.5,-0.45) {};
\node[selected,label=left:{$x_4$}] (x4) at (-1.15,-1.2) {};
\draw (c)--(t)--(z);
\draw (c)--(x1) (c)--(x2) (c)--(x3) (c)--(x4);
\path[use as bounding box] (-2.05,-1.6) rectangle (3.05,1.45);
\end{tikzpicture}

{\small (a) $G_1$}
\end{minipage}
\hfill
\begin{minipage}[t]{0.48\textwidth}
\centering
\begin{tikzpicture}[
  x=1cm,y=1cm,
  vertex/.style={circle,draw=black,fill=white,inner sep=0pt,minimum size=5pt},
  every node/.style={font=\small},
  every path/.style={line width=0.7pt}
]
\node[vertex,label=above:{$u$}] (u) at (0,0) {};
\node[vertex,label=above:{$v$}] (v) at (1.4,0) {};
\node[vertex] (u1) at (-1.3,1.0) {};
\node[vertex] (u2) at (-1.5,0) {};
\node[vertex] (u3) at (-1.3,-1.0) {};
\node[vertex] (v1) at (2.7,0.85) {};
\node[vertex] (v2) at (2.7,-0.85) {};
\draw (u)--(v);
\draw (u)--(u1) (u)--(u2) (u)--(u3);
\draw (v)--(v1) (v)--(v2);
\path[use as bounding box] (-1.85,-1.6) rectangle (3.25,1.45);
\end{tikzpicture}

{\small (b) $G_2$}
\end{minipage}
\caption{The trees $G_1$ and $G_2$ in Example~\ref{ex:separation}.}
\label{fig:separation}
\end{figure}

Both outer graphs are trees of order seven. Each has a matching of size two and a vertex cover of size two. In $G_1$, the edges $cx_1$ and $tz$ form a matching, and $\{c,t\}$ is a vertex cover. In $G_2$, one leaf edge incident with each center forms a matching, and $\{u,v\}$ is a vertex cover. Since matching size is at most fractional matching weight, which is at most vertex cover size, this proves $\nuf(G_1)=\nuf(G_2)=2$. Hence $\dc(G_1)=\dc(G_2)=3$.

Because $\dc(P_3)=1$, computing $D(G_i,P_3)$ amounts to maximizing $|A|-3|N_{G_i}(A)|$ over independent sets $A$. For $G_1$, the independent set $A$ indicated in Figure~\ref{fig:separation}(a) gives $|A|-3|N_{G_1}(A)|=1$. We check that no larger value occurs. If $c\in A$, then $|A|\le2$ and $|N_{G_1}(A)|\ge5$. If $t\in A$, then $|A|\le5$ and $|N_{G_1}(A)|\ge2$. Otherwise $A$ consists of some $x_i$ and possibly $z$. Without $z$, a nonempty such set has at most four vertices and satisfies $|N_{G_1}(A)|=1$. With $z$ alone it has value $-2$, and with $z$ and at least one $x_i$ it has at most five vertices and satisfies $|N_{G_1}(A)|=2$. These cases give $D(G_1,P_3)=1$.

For $G_2$, first suppose that $A$ contains neither center. A nonempty set of leaves on one side has at most three vertices and satisfies $|N_{G_2}(A)|=1$. A set meeting both sides has at most five vertices and satisfies $|N_{G_2}(A)|=2$. If $u\in A$, then $|A|\le3$ and $|N_{G_2}(A)|\ge4$. If $v\in A$, then $|A|\le4$ and $|N_{G_2}(A)|\ge3$. Every case gives $|A|-3|N_{G_2}(A)|\le0$, and the empty set attains zero. Thus $D(G_2,P_3)=0$.

The exact values are therefore $\nuf(G_1\circ P_3)=10$ and $\nuf(G_2\circ P_3)=21/2$. By Proposition~\ref{prop:reduction}, these are also the respective fractional monitoring numbers. In contrast, the lower bound~\eqref{eq:ma} gives nine for both products. This pair shows that the orders and fractional matching numbers of the factors do not determine the product value.
\end{example}

\FloatBarrier
\section{Paths, stars and further consequences}\label{sec:examples}

For a path of order $r$, its fractional matching number equals its matching number, namely $\lfloor r/2\rfloor$, since a matching and a vertex cover of that size exist. Thus $\dc(P_r)=0$ when $r$ is even and $\dc(P_r)=1$ when $r$ is odd.

\begin{proposition}\label{prop:paths}
For all integers $m,n\ge2$, we have $\demf(P_m\circ P_n)=mn/2$ and $\dem(P_m\circ P_n)=mn-\lceil m/2\rceil\lceil n/2\rceil$.
\end{proposition}

\begin{proof}
Every nonempty independent set $A$ of $P_m$ satisfies $|A|\le2|N_{P_m}(A)|$. Indeed, each vertex of $A$ has a neighbor and each vertex of its neighborhood is adjacent to at most two vertices of $A$. If $n$ is even, then $\dc(P_n)=0$. If $n$ is odd, then $n\ge3$ and $\dc(P_n)=1$, so $\dc(P_n)|A|\le2|N_{P_m}(A)|\le n|N_{P_m}(A)|$. Theorem~\ref{thm:main} gives the fractional formula. The integral formula follows from $\alpha(P_r)=\lceil r/2\rceil$ and the integral identity recalled after Proposition~\ref{prop:reduction}.
\end{proof}

\begin{example}\label{ex:stars}
Let $G=K_{1,r}$ and $H=K_{1,s}$, where $r,s\ge1$. The leaves of $H$ give $\dc(H)=s-1$. A nonempty independent set of the outer star is either its center or a subset of its leaves. The center gives a nonpositive value in~\eqref{eq:D}, and among subsets of leaves the maximum is attained by all $r$ leaves. Therefore $D(K_{1,r},K_{1,s})=\max\{0,r(s-1)-(s+1)\}$, and
\[
\demf(K_{1,r}\circ K_{1,s})
=\frac{(r+1)(s+1)-\max\{0,r(s-1)-(s+1)\}}2.
\]
For example, $\demf(K_{1,3}\circ K_{1,3})=7<8$. Thus connectedness of both factors does not force equality in the upper bound.
\end{example}

\begin{remark}\label{rem:computation}
The formula can be evaluated by standard linear programming. Construct $G\circ H$, compute $\nuf(G\circ H)$ using fractional matching~\cite{ScheinermanUllman1997}, and recover $D(G,H)=mn-2\nuf(G\circ H)$. This takes polynomial time for explicitly given factors. The purpose of~\eqref{eq:D} is to expose the independent set neighborhoods that determine the value.
\end{remark}

\section{Concluding remarks}

The fractional monitoring problem on a lexicographic product reduces to fractional matching through the established endpoint monitoring property. Combining the classical critical difference identity with the neighborhood decomposition by fibers yields the exact formula in Theorem~\ref{thm:main}. This expresses the contribution of the inner factor through a single numerical parameter and the contribution of the outer factor through the neighborhoods of its independent sets.

The formula gives a necessary and sufficient condition for equality in the fractional-matching lower bound of Ma et al.~\cite{MaChengMaoWang2022}, under the stated connectedness assumption on the outer factor, and characterizes the outer graphs attaining the upper bound for every inner graph. The two trees in Example~\ref{ex:separation} show that replacing the outer neighborhood information by its fractional matching number loses information needed for the exact product value. Further structural questions include explicit formulas for additional outer graph classes and a characterization of the products satisfying $\dem(G\circ H)=\demf(G\circ H)$.

\section*{Funding}
This work was supported in part by the National Natural Science Foundation of China under Grant 12471329, and in part by the National Key Research and Development Program of China under Grant 2023YFA1009604.

\section*{Data availability}
No external datasets are used. The graphs and constructions needed for the results are specified in the article.

\section*{Declaration of generative AI and AI-assisted technologies in the manuscript preparation process}
During the revision of this manuscript, the authors used ChatGPT to
assist with language and \LaTeX{} editing, to improve the exposition, and
to identify possible errors and explore possible simplifications and
refinements of the arguments. All AI-assisted suggestions incorporated into the manuscript
were independently assessed, reconstructed, and verified by the authors,
who assume full responsibility
for the final content of the article.

\end{document}